\documentclass[]{spie}  

\usepackage{amsmath,amsfonts,amssymb}
\usepackage{graphicx}
\usepackage[colorlinks=true, allcolors=blue]{hyperref}

\title{In-orbit Commissioning of the Near-Infrared Spectrograph on the James Webb Space Telescope}

\author[a]{Torsten B\"oker}
\author[b]{Yasin Abul-Huda}
\author[j]{Martin Altenburg}
\author[c]{Catarina Alves de Oliveira}
\author[b]{Katie Bechtold}
\author[b]{Tracy Beck}
\author[a]{Stephan M. Birkmann}
\author[k]{Nina Bonaventura}
\author[j]{Ralf Ehrenwinkler}
\author[c]{Pierre Ferruit}
\author[h]{David E. Franz}
\author[i]{Giovanna Giardino}
\author[k]{Peter Jakobsen}
\author[f]{Peter Jensen}
\author[f]{Delphine Jollet}
\author[b]{Diane Karakla}
\author[j]{Hermann Karl}
\author[b]{Charles Keyes}
\author[d]{Nimisha Kumari}
\author[h]{Matthew Lander}
\author[e]{Marcos L\'{o}pez-Caniego}
\author[a]{Nora L\"{u}tzgendorf}
\author[d]{Elena Manjavacas}
\author[c]{Anthony Marston}
\author[j]{Marc Maschmann}
\author[j]{Peter Mosner}
\author[b]{James Muzerolle}
\author[b]{Patrick Ogle}
\author[b]{Maria Pena Guerrero}
\author[b]{Rachel Plesha}
\author[b]{Charles Proffitt}
\author[h]{Robert Rapp}
\author[a]{Timothy Rawle}
\author[g]{Bruno Rodr\'iguez Del Pino}
\author[b]{Elena Sabbi}
\author[j]{Arne Sauer}
\author[a]{Marco Sirianni}
\author[h]{Corbett Smith}
\author[a]{Maurice te Plate}
\author[b]{Glenn Wahlgren}
\author[b]{Emily Wislowski}
\author[b]{Rai Wu}
\author[d]{Peter Zeidler}
\author[h]{Christian A. Zincke}

\affil[a]{European Space Agency, Space Telescope Science Institute, Baltimore, Maryland, USA}
\affil[b]{Space Telescope Science Institute, Baltimore, Maryland, USA}
\affil[c]{European Space Agency, ESAC, Madrid, Spain}
\affil[d]{AURA for the European Space Agency, STScI, Baltimore, Maryland, USA}
\affil[e]{Aurora Technology for the European Space Agency, ESAC, Madrid, Spain}
\affil[f]{European Space Agency, ESTEC, Noordwijk, The Netherlands}
\affil[g]{Centro de Astrobiolog\'ia, CSIC-INTA, Madrid, Spain}
\affil[h]{NASA Goddard Space Flight Center, Greenbelt, Maryland, USA}
\affil[i]{ATG Europe for the European Space Agency, Noordwijk, The Netherlands}
\affil[j]{Airbus Defence and Space GmbH, Ottobrunn, Germany}
\affil[k]{Cosmic Dawn Center, Niels Bohr Institute, University of Copenhagen, Denmark}

\authorinfo{Further author information: Torsten.Boeker@esa.int}

\newcommand{\mum}{\,{\mu {\rm m}}}

\newcommand{\am}{^{\prime}}
\newcommand{\bdm}{\begin{displaymath}}
\newcommand{\edm}{\end{displaymath}}
\newcommand{\beq}{\begin{equation}}
\newcommand{\eeq}{\end{equation}}
\newcommand{\bit}{\begin{itemize}}
\newcommand{\eit}{\end{itemize}}
\newcommand{\ben}{\begin{enumerate}}
\newcommand{\een}{\end{enumerate}}
\newcommand{\bfi}{\begin{figure}[htb]}
\newcommand{\bpfi}{\begin{figure}[p]}

\begin{document} 
\maketitle

\begin{abstract}
The Near-Infrared Spectrograph (NIRSpec) is one of the four focal plane instruments on the James Webb Space Telescope which was launched on Dec. 25, 2021. We present an overview of the as-run NIRSpec commissioning campaign, with particular emphasis on the sequence of activities that led to the verification of all hardware components of NIRSpec. We also discuss the mechanical, thermal, and operational performance of NIRSpec, as well as the readiness of all NIRSpec observing modes for use in the upcoming JWST science program.  
\end{abstract}


\section{INTRODUCTION}
\label{sec:intro}  
The launch of the James Webb Space Telescope (JWST) on Dec. 25, 2021 marked the start of a roughly six-months long commissioning campaign aimed at deploying and aligning all observatory components, as well as characterizing the performance of the optical telescope element (OTE) and the science instruments (SIs). One of these science instruments is the Near-Infrared Spectrograph (NIRSpec), which was designed and built for the European Space Agency (ESA) by Airbus Defense and Space GmbH in Ottobrunn, Germany.

NIRSpec is a versatile near-infrared spectrograph with a variety of observing modes. It has a large field of view ($\approx 3\am \times 3\am$) and is highly sensitive over the wavelength range $0.6 - 5.3\mum$. The main purpose of NIRSpec is to enable low-, medium-, and high-resolution ($\frac{\lambda}{\Delta\lambda}=100$, 1000, and 2700, respectively) near-infrared spectroscopy in support of all JWST science themes. In its multi-object spectroscopy mode, NIRSpec can observe up to 200 astronomical objects simultaneously. A detailed description of the NIRSpec design, observing modes, and scientific use cases can be found in [\citenum{Jakobsen22,Ferruit22,Boeker22,Birkmann22}].

The objectives of the NIRSpec commissioning campaign, and the planned sequence of activities to achieve them have been described previously [\citenum{Boeker16}]. Here, we provide an overview of the as-run timeline and some results measured from in-orbit data.

\section{Timeline of Events and Milestones}
Commissioning an observatory as complex as JWST involves a large amount of coordination and dependencies between the various subsystems. Critical milestones such as the launch itself, the numerous deployment activities during the transfer to L2, the alignment of the OTE, and verification of the thermal performance of the sunshield were all successfully completed, but are not discussed further in this paper.

\begin{figure} [ht]
   \begin{center}
   \begin{tabular}{c} 
   \includegraphics[width=0.9\textwidth]{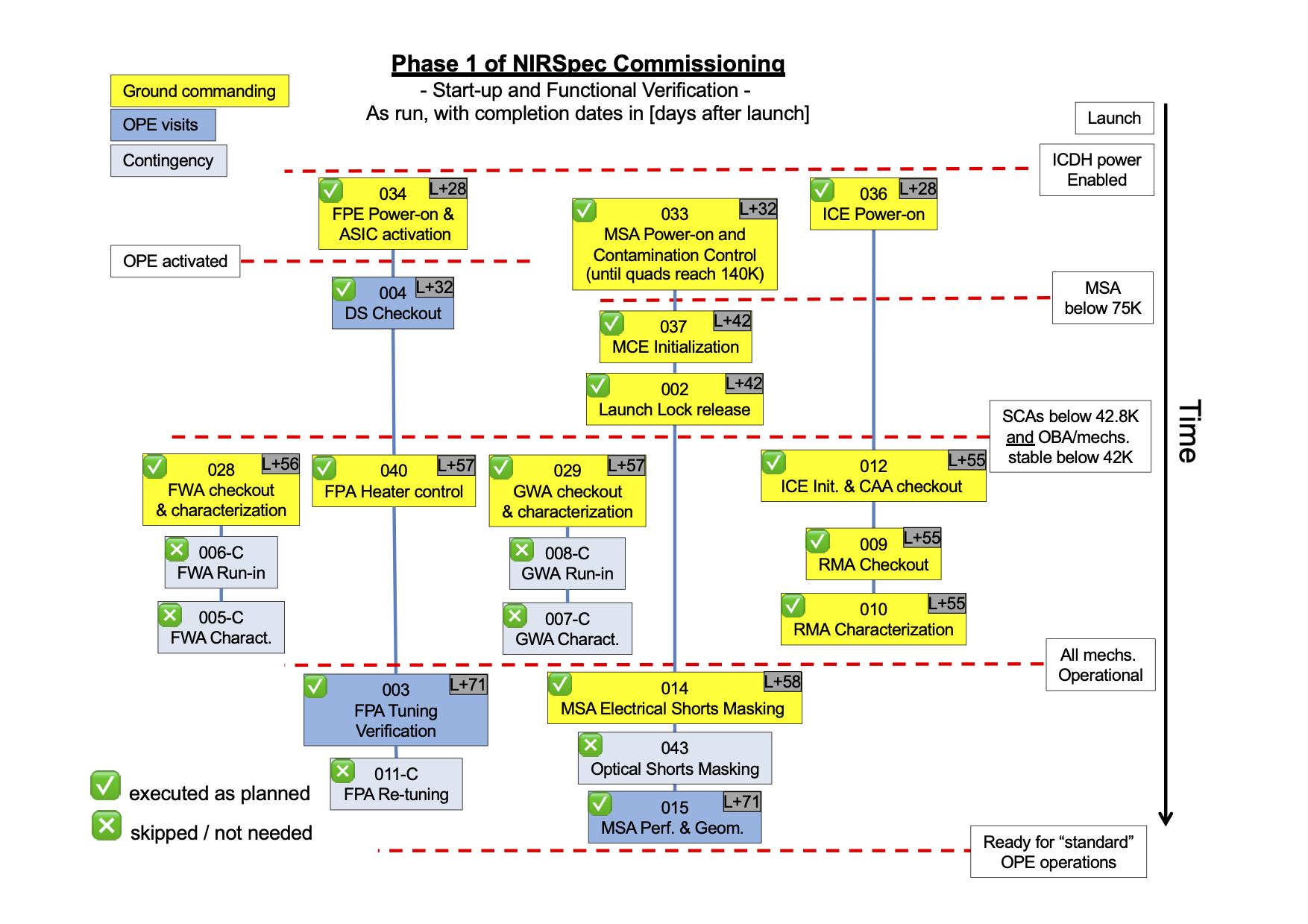}
   \end{tabular}
   \end{center}
   \caption[example]{\label{fig:phase1} 
Activity flow for the first phase of NIRSpec commissioning, focused on the power-on and functional verification of all electronics and mechanisms. Execution dates relative to launch are listed in the upper right corner of each box. Activities in yellow were executed via real-time commanding, while activities in blue were run via the autonomous on-board software (Observation Plan Executive, OPE). None of the planned contingency activities (light blue boxes) were needed.}
   \end{figure}

\begin{figure} [ht]
   \begin{center}
   \begin{tabular}{c} 
   \includegraphics[width=0.9\textwidth]{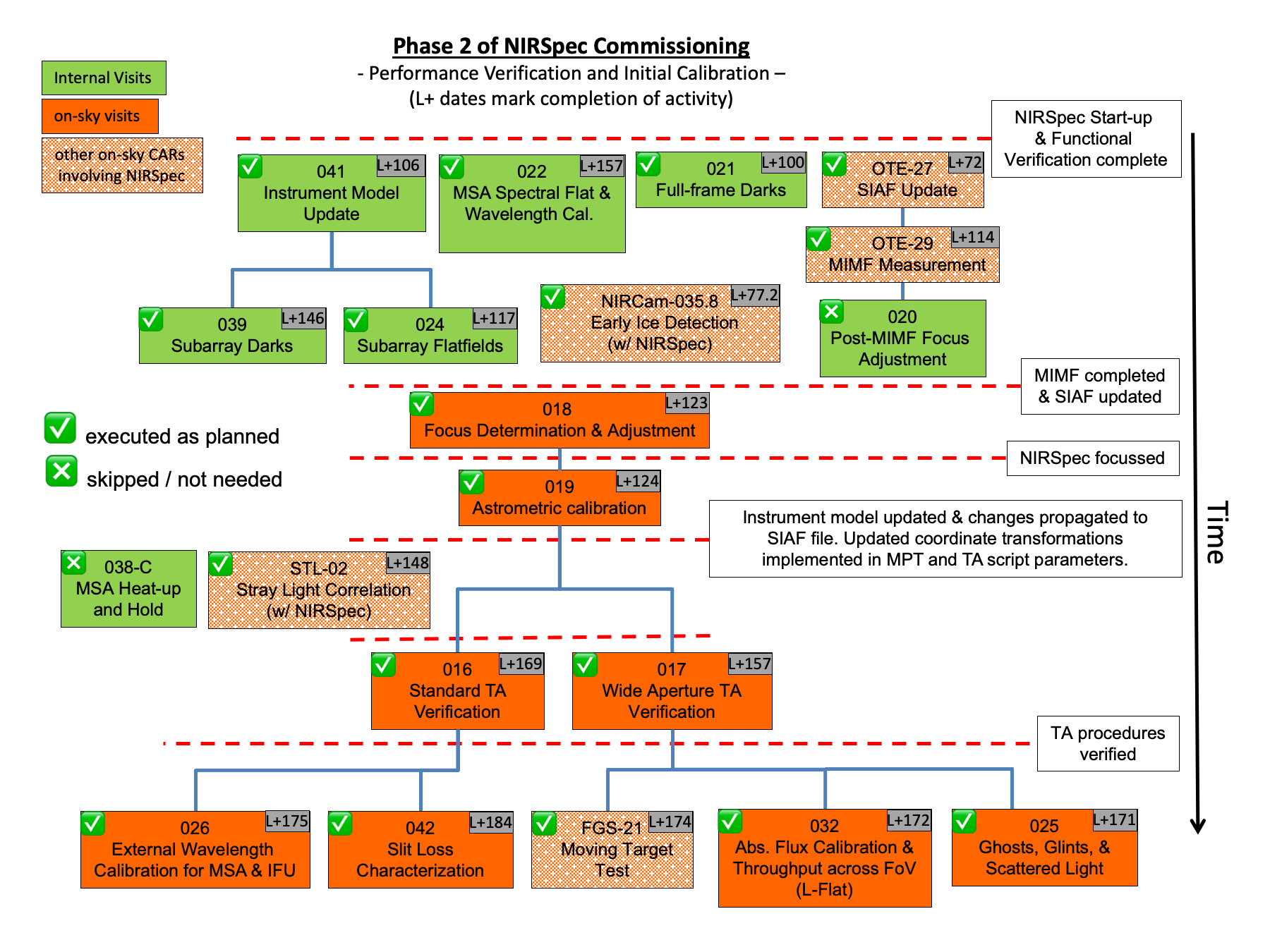}
   \end{tabular}
   \end{center}
   \caption[example]{\label{fig:phase2} 
Activity flow for the second phase of NIRSpec commissioning, focused on the performance verification and initial calibration of all NIRSpec science modes. A number of activities were split up in multiple parts, in which case the dates listed in the upper right corner of each box mark the completion date, i.e. the execution of the last program portion. Green boxes denote internal exposures (which could be started before telescope alignment was complete), while orange boxes mark on-sky observations requiring a fully aligned telescope. Light orange boxes denote observatory-level commissioning activities that involve the use of NIRSpec, but are not led directly by the NIRSpec commissioning team.}
   \end{figure}

Instead, we will focus only on those events and milestones that directly affect the NIRSpec commissioning timeline. Figures~\ref{fig:phase1} and \ref{fig:phase2} summarize the 33 dedicated NIRSpec Commissioning Activity Requests (CARs) that were run during the campaign. These had been extensively reviewed and optimized for efficiency prior to launch, and thus represent the minimal set of activities needed to enable the scientific use of NIRSpec and all its observing modes. 

It is notable that the flow depicted in Figs.~\ref{fig:phase1} and \ref{fig:phase2} differs only minimally from the original plan devised years ago [\citenum{Boeker16}]. In particular, none of the planned contingency activities (such as mechanism run-ins or re-tuning of the detector readout circuitry) were necessary. Given the long hiatus since the last cryogenic use of the instrument during the OTIS test campaign [\citenum{Kimble18}] in 2017, this is  testament to the careful handling of the JWST hardware during integration, testing, and transport.

The major milestones for NIRSpec commissioning (indicated on the right in Figs.~1 and 2) were the following:

\begin{enumerate}
    \item The successful execution of the onboard script that controls and maintains the temperature of the microshutter quadrants 10-15\,K above the ISIM environment until the latter dropped below 140K, in order to avoid any risk of shutter contamination by water ice or other volatiles.
    
    \item The NIRSpec  Optical Bench Assembly (OBA) reaching sufficiently cold temperatures for the safe operation of the various mechanisms (MSA magnet arm, grating and filter wheel, re-focus mechanism), and the active control of the Focal Plane Assembly (FPA) temperature.
    
    \item The verification of the mechanical and operational performance of these mechanisms and all lamps in the Calibration Assembly (CAA), which was completed on day L+57d.
    
    \item The detailed characterization of detector performance (dark current, read noise, and bad pixel maps) and functionality of the $\approx 250000$ individual shutters of the MSA. The results, which are detailed in companion papers [\citenum{Birkmann22b,Rawle22}], confirmed the expected in-flight performance, and marked the start of Phase 2 of NIRSpec commissioning, i.e. the autonomous operation via the Observation Plan Executive (OPE) which obtains science data without real-time intervention.
    
    \item The completion of the OTE alignment on day L+120d enabled the first on-sky NIRSpec observations, in particular the `focus sweep' which established the optimal positioning of the internal focus adjustment mechanism (RMA), and allowed a first assessment of the image quality and total wavefront error across the NIRSpec field of view.
    
    \item We then obtained images of a dense star field for which highly accurate measurements of the stellar positions exist. These observations of the `astrometric reference field' allowed us to precisely derive the magnification and distortions of the NIRSpec optical train, which is captured in the parametric instrument model described in [\citenum{Luetzgendorf22}]. The resulting coordinate transformation between detector, MSA, and sky are necessary to be able to position science targets in the various NIRSpec apertures.
    
    \item Once the accuracy of the instrument model was confirmed via a series of successful target acquisitions, the last portion of the NIRSpec commissioning campaign obtained on-sky observations of stars and planetary nebulae for flux and wavelength calibration, which enabled the creation of various `reference files' and other products needed to deliver calibrated NIRSpec spectra via the data reduction pipeline.
    
\end{enumerate}
In what follows, we will describe these milestones and the resulting performance metrics in more detail.

\section{NIRSpec Hardware Performance}

\subsection{Cooldown Profile and Temperature Stability}

\begin{figure} [h]
   \begin{center}
   \begin{tabular}{c} 
   \includegraphics[width=0.9\textwidth]{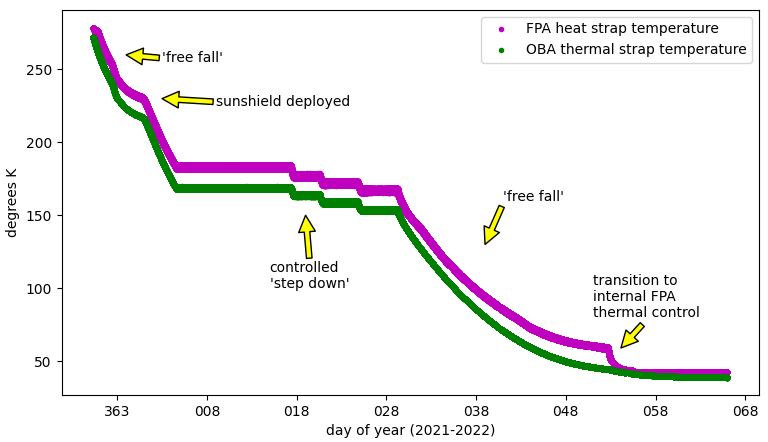}
   \end{tabular}
   \end{center}
   \caption[example]{\label{fig:cooldown} 
Cooldown profile for the NIRSpec focal plane array (FPA, purple) and optical bench assembly (OBA, green). 
It took about 58 days for the NIRSpec detectors to reach and stabilize at their operating temperature of 42.8\,K, which enabled the start of the Performance Verification phase of NIRSpec commissioning.  }
   \end{figure}

The temperature profiles of the NIRSpec focal plane array and the optical bench assembly are shown in Fig.~\ref{fig:cooldown}. Throughout the roughly two months long cooling phase, and using heaters on the heat strap connecting the FPA to its dedicated radiator, the detectors were kept 10-12\,K warmer than the NIRSpec bench, in order to guard against condensation of water ice and other contaminants on the detectors. A similar approach was taken for the microshutter assembly (not shown). Once the external FPA heaters were disabled on day L+54d, the detectors rapidly dropped to their final operating temperature of $\approx 42.8$\,K. This temperature is now maintained to within 10\,mK by dedicated stability heaters on the FPA itself, thus ensuring stable detector performance for all science modes. This milestone marked the start of the Performance Verification phase of NIRSpec commissioning.

\subsection{Mechanism Performance}
During the roughly 2 month long cooldown phase, the NIRSpec commissioning activities were limited to powering on the three electronics boxes (ICE, FPE, and MCE), characterizing the mechanisms (GWA, FWA, RMA), and verifying the functionality of the internal calibration lamp assembly. Despite the long hiatus since the last cryogenic use of NIRSpec during the OTIS test campaign in 2017, all mechanisms behaved well within their performance specifications, and none of the planned contingency activities such as wheel run-ins were required during commissioning. In particular, the measured torque values for the GWA and FWA were very close to those measured during previous ground-based cryo-campaigns.

This is important, because it allowed a quick and accurate calibration of the position sensors on the GWA. As described in more detail in [\citenum{Alves22}], this is an essential pre-requisite for the precision of the NIRSpec parametric instrument model that informs the coordinate transformations between sky, MSA, and FPA used during target acquisition, as well as for the extraction and calibration of NIRSpec spectra. 

\subsection{Detector Performance}
As soon as the NIRSpec focal plane array was put under active temperature control, and had stabilized at its operating temperature of 42.8K, dedicated detector characterization activities were executed to verify the ASIC tuning parameters, bias values, and the in-orbit readout noise and dark current behavior. The detailed results of these activities are described in [\citenum{Birkmann22b}].

In summary, all performance metrics of the detectors were as expected, which removed any need to execute a re-tuning of the ASIC parameters. In particular, the measured total readout noise is fully in line with pre-flight expectations for both NIRSpec readout modes (traditional and IRS2, see [\citenum{Birkmann22b}] for details). Compared to pre-flight ground testing, a slight elevation of the dark current was observed which most likely is due to residual signal from low-level cosmic rays that the pipeline does not fully correct for. This does not, however, contribute significantly to the total noise, and therefore does not impact the NIRSpec sensitivity which is read noise-limited for most observations (except for very high backgrounds or very bright targets).

\subsection{Performance of the Micro-Shutter Assembly}
A long-standing concern for the NIRSpec in-orbit performance has been the number of failed microshutters in the MSA, which tends to increase with thermal cycling and/or exposure to vibration and acoustic loads. Given that the last cryogenic performance test of the MSA occurred in 2017, it was a major relief to see that neither the intermittent integration activities, the transport to the launch site, nor the launch itself had any significant effect on the failed shutter statistics. This demonstrates the success of the extra measures taken by NASA to protect this sensitive subsystem.

The detailed evolution of the failed shutter numbers, and a map of their in-orbit distribution are presented in [\citenum{Rawle22}]. Here, we only summarize the evolution of the shutter statistics since the start of ISIM-level testing in Table\,\ref{tab:msa}.

\begin{table}[ht]
\caption{Evolution of the failed shutter statistics of the NIRSpec Microshutter Assembly} 
\label{tab:msa}
\begin{center}       
\begin{tabular}{|l|c|c|c|c|} 
\hline
\rule[-1ex]{0pt}{3.5ex}  & End of CV3 & End of OTIS & First Light & End of Comm. \\
                        & (Feb. 2016) & (Sept. 2017) & (Feb. 2022) & (June 2022) \\
\hline
\rule[-1ex]{0pt}{3.5ex}  {\bf Total $\#$ of shutters}  & {\bf 249660}
 & {\bf 249660} & {\bf 249660} & {\bf 249660}  \\
\hline
\rule[-1ex]{0pt}{3.5ex}  Vignetted  & 23705 & 23705 & 24024 &  24024  \\
\hline 
\rule[-1ex]{0pt}{3.5ex}  Failed Open & 19 & 18 & 20 & 22  \\
\hline
\rule[-1ex]{0pt}{3.5ex}  Failed Closed & 13178 & 13425 & 15335 & 15893  \\
\hline
\rule[-1ex]{0pt}{3.5ex}  Masked & 18123 & 20450 & 20924 & 23628  \\
\hline 
\rule[-1ex]{0pt}{3.5ex}  {\bf Usable $\#$ of shutters} & 194654 & 192080 & 189377 & 186115  \\
{\bf (Percentage)} & {\bf (86.1\%)}  & {\bf (85.0\%)} & {\bf (83.9\%)} & {\bf (82.5\%)} \\
\hline 
\end{tabular}
\end{center}
\end{table}

\subsection{NIRSpec Image Quality and Impact on Slit Losses}\label{sec:wfe}
The in-orbit optical performance of the JWST telescope is significantly better than expected, with lower wavefront errors and therefore a sharper point spread function (PSF) across the entire field of view [\citenum{Feinberg22}]. More specifically, the JWST PSF is diffraction limited already at $1.1\mum$, and reaches a Strehl ratio of 0.9 at $4\mum$ across the entire field of view.

\begin{figure} [ht]
   \begin{center}
   \begin{tabular}{c} 
   \includegraphics[width=0.9\textwidth]{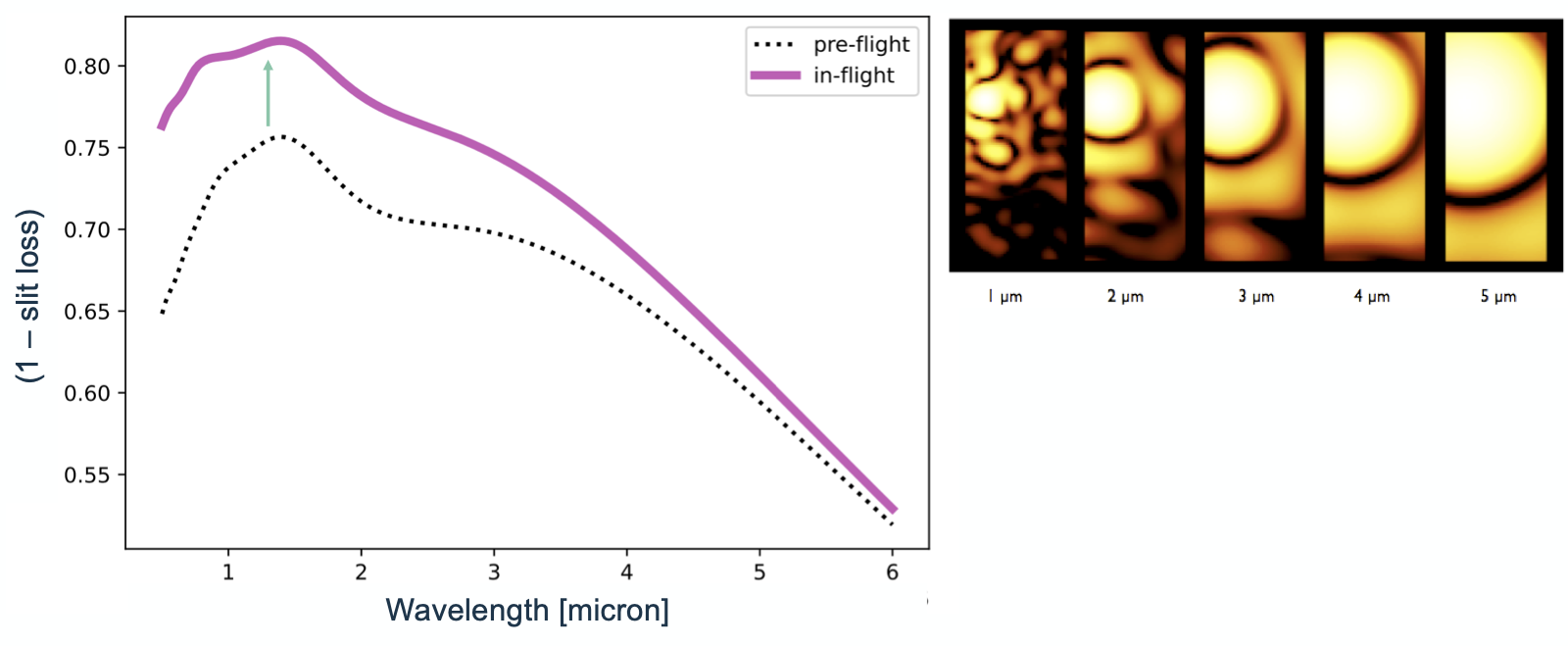}
   \end{tabular}
   \end{center}
   \caption[example]{\label{fig:losses} 
Left: the measured in-orbit slit throughput (1 - losses, purple curve) is compared to the pre-flight expectation (dotted curve) based on the simulations shown in the right panel. Right: illustration of `slit losses' caused by the physical truncation of the NIRSpec PSF at the microshutter (or 200\,mas wide fixed slit) aperture.  }
   \end{figure}

While this is clearly good news for all JWST instruments, the NIRSpec sensitivity stands to gain the most. This is because a narrower PSF minimizes `aperture losses' caused by the physical truncation of the point spread function by the narrow apertures of the NIRSpec fixed slits and microshutters. This is illustrated in the right panel of Fig.\,\ref{fig:losses}, which shows PSF simulations based on pre-flight expectations for the OTE wavefront error, compared to the physical size of a microshutter. From such simulations, the expected aperture throughput (1 - slit loss) can be computed as a function of wavelength (left panel, dotted curve), and compared to the in-flight measurements (solid purple curve). As the comparison shows, the NIRSpec slitlosses are significantly smaller than expected at all wavelengths, with peak gains of more than 10\% at the shorter wavelengths.
The resulting benefit to the overall NIRSpec sensitivity is discussed in Section\,\ref{sec:sensitivity}

\section{NIRSpec Science Performance}

\subsection{Target Acquisition Accuracy}
The major challenge in enabling NIRSpec's multi-object spectroscopy (MOS) mode is to precisely align the science targets in their dedicated microshutters. Since the microshutters are only about 200 milli-arcsec wide, the stars need to be placed with an accuracy of 20 milli-arcsec or better. This was a major hurdle on the path to commissioning the NIRSpec instrument, and required precise tuning of the parametric instrument model [\citenum{Luetzgendorf22}] to derive the coordinate transformations between the detector, MSA, and sky planes.

Once these transformations are in hand, they are uploaded for use by the JWST onboard software, which needs to autonomously perform a suite of rather complex calculations (see [\citenum{Keyes18}] for details):
\begin{enumerate}
\item{measure the precise locations of various reference stars on the NIRSpec detector from the initial target acquisition exposures}

\item{from these measurements and the known celestial positions of the reference stars, calculate the actual observatory pointing (RA, Dec) and roll angle. }

\item{compute the corrective SAM (small angle maneuver, which includes a roll angle adjustment) required for the science targets to fall into their dedicated slitlets on the microshutter array, and instruct the Attitude Control System to execute it.}
\end{enumerate}

\begin{figure} [ht]
   \begin{center}
   \begin{tabular}{c} 
   \includegraphics[width=0.9\textwidth]{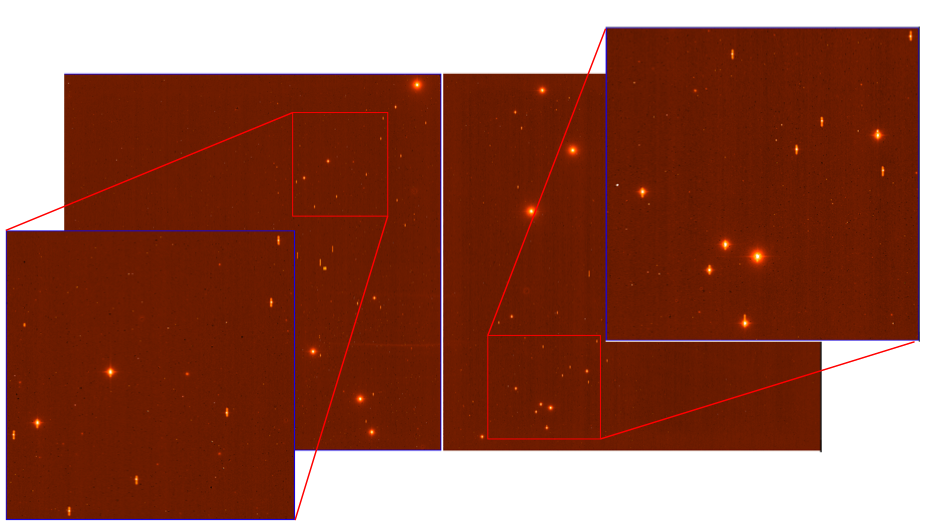}
   \end{tabular}
   \end{center}
   \caption[example]{\label{fig:ta} 
NIRSpec target acquisition `confirmation' image, obtained after execution of the corrective small angle maneuver that adjusts the telescope pointing and roll angle such that all science targets fall into their intended microshutter slitlets.}
   \end{figure}

After completion of the corrective SAM, all science targets should be positioned near the center of their dedicated `slitlets' comprised of three open microshutters aligned in the cross-dispersion direction. After fixing a few minor bugs in the onboard computations, the end-to-end process for MOS target acquisition was successfully demonstrated on June 22, 2022. Figure\,\ref{fig:ta} shows a ‘confirmation image’ obtained after the execution of the corrective SAM, through the science filter, and still with the grating wheel in the MIRROR position to allow for an undispersed image. As can be seen most clearly in the enlarged cutouts, each of the opened tiny ($1 \times 3$) shutter slitlets contains a nicely centered star.

A number of MOS visits for Cycle 1 science programs have been successfully executed since this milestone, and early analysis of the data indicates a typical placement accuracy of 8 milli-arcsec per axis, which is well within the required tolerances.

\subsection{Optical Throughput and Sensitivity}\label{sec:sensitivity}
\begin{figure} [ht]
   \begin{center}
   \begin{tabular}{c} 
   \includegraphics[width=0.9\textwidth]{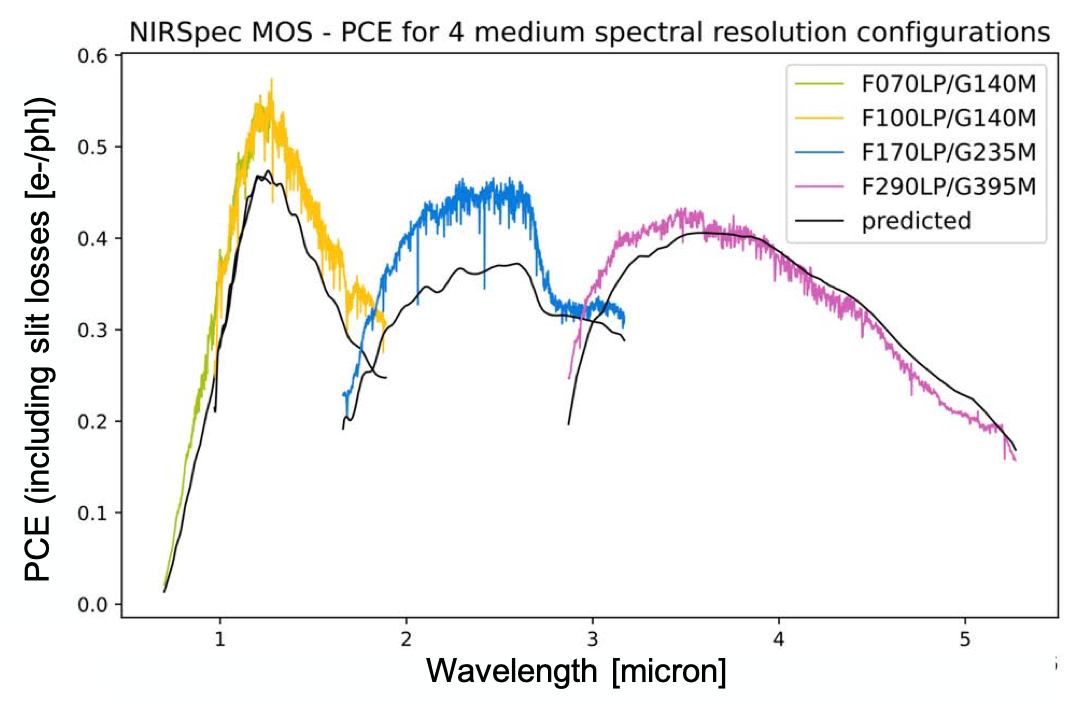}
   \end{tabular}
   \end{center}
   \caption[example]{\label{fig:pce} 
Measured photon conversion efficiency (PCE) for the NIRSpec medium-spectral resolution configurations in NIRSpec MOS mode (or the 200mas wide fixed slits). The black, solid lines correspond to the predicted PCEs (using a pre-launch model of the telescope PSF and the associated slit loss estimates). The colored curves show the values measured in flight. Any spikes or other narrow features in the measured PCE (either positive or negative) are due to residuals from bad detector pixels and/or mismatched absorption lines in the standard star spectrum used for this measurement. These artefacts will be removed in the final processing.}
   \end{figure}

As mentioned in Section\,\ref{sec:wfe}, the better than expected optical quality of the JWST telescope has a very beneficial impact on the NIRSpec performance because of the reduced losses at the slit apertures. As an example, we show in Fig.\,\ref{fig:pce} the photon conversion efficiency (PCE) as a function of wavelength for observations with the medium-resolution gratings, and through a 200\,mas wide aperture (i.e. through the microshutters or the S200 fixed slits). The significant gain in PCE relative to the pre-launch expectations (black lines) for most wavelengths is obvious, and exceeds 10\% for $\lambda \lesssim 3\mum$

As discussed in more detail in [\citenum{Giardino22}], the NIRSpec PCE generally exceeds the pre-launch expectations. For the fixed-slit and multi-object spectroscopy modes, lower than predicted PCEs are only observed at the red end of some spectral configurations, and always remain within 10\% of the predictions. The integral-field spectroscopy mode shows more contrasted results, with relative differences reaching -20\% above $4\mum$ but often exceeding +30\% below $2\mum$. These differences can have multiple origins, but mainly reflect the limited accuracy of the throughput/reflectivity measurements  conducted on the ground at component- or instrument-level, as well as the difficulty to accurately predict the aperture and diffraction losses (see also Sections 3.1 and 3.2 of [\citenum{Jakobsen22}]).

\section{Science Readiness of the NIRSpec Observing Modes}
Verifying the functionality of the NIRSpec hardware is necessary, but not sufficient for executing general observer science programs. As described in more detail in [\citenum{kimble22}], an end-to-end verification of planning, executing, downlinking, processing, and archiving science data involves many subsystems, both of the observatory and the ground system.

Towards the end of the commissioning campaign, the NIRSpec on-sky programs were executed with the same observing templates used by regular science programs, in order to verify the correct behavior of the observatory when identifying and guiding on guide stars, performing target acquisitions, tracking moving targets, or simply executing mosaics or dither patterns. The exposures from these programs were also used to evaluate the quality of the products from the STScI science data pipeline (and thus of the early calibration files delivered by the SIs).

Following the analysis of the data from these programs, the end game of commissioning then was to formally certify all 17 JWST observing modes as being “ready for Cycle 1 science.” For each mode, a variety of pre-defined quantitative performance metrics were reviewed by the relevant stakeholders in a dedicated meeting. For NIRSpec, there are four science modes:

\begin{enumerate}
    \item{Multi-Object Spectroscopy (MOS)} 
    \item{Integral-Field Spectroscopy (IFS)}
    \item{Fixed Slit Spectroscopy (FSS)}
    \item{Bright Object Time Series Observations (BOTS)}
\end{enumerate}

Given the outstanding performance of the NIRSpec hardware and the observatory in general, each of the four modes passed its respective readiness review without major issues. A couple of minor `liens' were agreed, one for the MOS mode (related to the need to carefully vet guide star candidates before attempting the target acquisition), and one for the IFS mode (related to a small offset in the actual placement of the aperture reference point compared to the one used by the planning system). Except in rare circumstances, neither prevents the execution of science programs, and both will be cleared soon by pending updates to the planning system.

In conclusion, the NIRSpec commissioning campaign has successfully verified the functionality of all NIRSpec hardware, and has demonstrated the science readiness of all its observing modes. This was also evident from the impressive quality of the Early Release Observations [\citenum{ponto22}] presented by NASA on July 12, 2022, which prominently featured science results from two NIRSpec modes (MOS and IFS).

\section{Summary}
The commissioning of JWST, and NIRSpec in particular, was a major effort, as illustrated by the following numbers: the NIRSpec commissioning team was comprised of 49 individuals, who supported a total of 186 days of around-the-clock monitoring of the NIRSpec status and telemetry, amounting to 1480 shifts of 8 hours each. A total of 33 NIRSpec activities were executed, during which 284 hours worth of NIRSpec exposures were acquired, for a combined 1.3\,TB of raw data. While the detailed analysis of the data and creation of the commissioning calibration reference files will continue for another few months, the early results have clearly demonstrated that the NIRSpec performance exceeds expectations in almost all aspects. All four NIRSpec science modes have been declared ready for Cycle 1 observations, and the first few weeks of NIRSpec science data have confirmed that its users can look forward to a wide range of revolutionary results over the coming years.  

\acknowledgments 
We would like to acknowledge the hard work and dedication of the NIRSpec commissioning team, the NIRSpec science readiness team, and everyone involved in the design, construction, and commissioning of NIRSpec and the entire JWST observatory. We would not have such a wonderful science mission without them.  

\bibliography{report} 
\bibliographystyle{spiebib} 

\end{document}